# Improving 3D convolutional neural network comprehensibility via interactive visualization of relevance maps: Evaluation in Alzheimer's disease

Martin Dyrba[1,*], Moritz Hanzig[1,2], Slawek Altenstein[3,4], Sebastian Bader[2], Tommaso Ballarini[5], Frederic Brosseron[5,6], Katharina Buerger[7,8], Daniel Cantré[9], Peter Dechent[10], Laura Dobisch[11], Emrah Düzel[11,12], Michael Ewers[7,8], Klaus Fliessbach[5,6], Wenzel Glanz[11], John-Dylan Haynes[13], Michael T. Heneka[5,6], Daniel Janowitz[8], Deniz B. Keles[14], Ingo Kilimann[1,15], Christoph Laske[16,17,18], Franziska Maier[19], Coraline D. Metzger[11,12,20], Matthias H. Munk[16,18,21], Robert Perneczky[7,22,23,24], Oliver Peters[3,14], Lukas Preis[3,14], Josef Priller[3,4,25], Boris Rauchmann[22], Nina Roy[5], Klaus Scheffler[26], Anja Schneider[5,6], Björn H. Schott[27,28,29], Annika Spottke[5,30], Eike J. Spruth[3,4], Marc-André Weber[9], Birgit Ertl-Wagner[31,32], Michael Wagner[5,6], Jens Wiltfang[27,28,33], Frank Jessen[5,19,34], Stefan J. Teipel[1,15], for the ADNI[†], AIBL[†], DELCODE[‡] study groups

[1] German Center for Neurodegenerative Diseases (DZNE), Rostock, Germany
[2] Institute of Visual and Analytic Computing, University of Rostock, Rostock, Germany
[3] German Center for Neurodegenerative Diseases (DZNE), Berlin, Germany
[4] Department of Psychiatry and Psychotherapy, Charité – Universitätsmedizin Berlin, Campus Charité Mitte, Berlin, Germany
[5] German Center for Neurodegenerative Diseases (DZNE), Bonn, Germany
[6] Department for Neurodegenerative Diseases and Geriatric Psychiatry, University Hospital Bonn, Bonn, Germany
[7] German Center for Neurodegenerative Diseases (DZNE), Munich, Germany
[8] Institute for Stroke and Dementia Research (ISD), University Hospital, Ludwig Maximilian University, Munich, Germany
[9] Institute of Diagnostic and Interventional Radiology, Pediatric Radiology and Neuroradiology, Rostock University Medical Center, Rostock, Germany
[10] MR-Research in Neurosciences, Department of Cognitive Neurology, Georg-August-University, Goettingen, Germany
[11] German Center for Neurodegenerative Diseases (DZNE), Magdeburg, Germany
[12] Institute of Cognitive Neurology and Dementia Research (IKND), Otto-von-Guericke University, Magdeburg, Germany
[13] Bernstein Center for Computational Neuroscience, Berlin, Germany
[14] Department of Psychiatry and Psychotherapy, Charité – Universitätsmedizin Berlin, Campus Benjamin Franklin, Berlin, Germany
[15] Department of Psychosomatic Medicine, Rostock University Medical Center, Rostock, Germany
[16] German Center for Neurodegenerative Diseases (DZNE), Tuebingen, Germany
[17] Section for Dementia Research, Hertie Institute for Clinical Brain Research, Tuebingen, Germany
[18] Department of Psychiatry and Psychotherapy, University of Tuebingen, Tuebingen, Germany
[19] Department of Psychiatry, Medical Faculty, University of Cologne, Cologne, Germany
[20] Department of Psychiatry and Psychotherapy, Otto-von-Guericke University, Magdeburg, Germany
[21] Systems Neurophysiology, Department of Biology, Darmstadt University of Technology, Darmstadt, Germany
[22] Department of Psychiatry and Psychotherapy, University Hospital, Ludwig Maximilian University, Munich, Germany
[23] Munich Cluster for Systems Neurology (SyNergy), Ludwig Maximilian University, Munich, Germany
[24] Ageing Epidemiology Research Unit (AGE), School of Public Health, Imperial College London, London, UK
[25] Department of Psychiatry and Psychotherapy, Klinikum rechts der Isar, Technical University Munich, Munich, Germany
[26] Department for Biomedical Magnetic Resonance, University of Tuebingen, Tuebingen, Germany
[27] German Center for Neurodegenerative Diseases (DZNE), Goettingen, Germany
[28] Department of Psychiatry and Psychotherapy, University Medical Center Goettingen, Goettingen, Germany
[29] Leibniz Institute for Neurobiology, Magdeburg, Germany
[30] Department of Neurology, University Hospital Bonn, Bonn, Germany
[31] Institute for Clinical Radiology, Ludwig Maximilian University, Munich, Germany
[32] Department of Medical Imaging, University of Toronto, Toronto, Canada
[33] Neurosciences and Signaling Group, Institute of Biomedicine (iBiMED), Department of Medical Sciences, University of Aveiro, Aveiro, Portugal
[34] Excellence Cluster on Cellular Stress Responses in Aging-Associated Diseases (CECAD), University of Cologne, Cologne, Germany
* Corresponding author. Tel. +49-381-494-9482; Fax: +49-381-9472; mail: martin.dyrba@dzne.de

† Data used in preparation of this article were obtained from the Alzheimer's Disease Neuroimaging Initiative (ADNI) database (adni.loni.usc.edu) and the Australian Imaging Biomarkers and Lifestyle Study of Aging (AIBL) database (aibl.csiro.au). As such, the investigators within the ADNI and AIBL contributed to the design and implementation of ADNI and AIBL and/or provided data but did not participate in analysis or writing of this report. A complete listing of ADNI investigators can be found at http://adni.loni.usc.edu/wp-content/uploads/how_to_apply/ADNI_Acknowledgement_List.pdf. AIBL researchers are listed at aibl.csiro.au.

‡ Data used in this article were provided by the DELCODE study group of the Clinical Research of the German Center for Neurodegenerative Diseases (DZNE). Details can be found at www.dzne.de/en/research/studies/clinical-studies/delcode.

# Abstract

**Background:** Although convolutional neural networks (CNN) achieve high diagnostic accuracy for detecting Alzheimer's disease (AD) dementia based on magnetic resonance imaging (MRI) scans, they are not yet applied in clinical routine. One important reason for this is a lack of model comprehensibility. Recently developed visualization methods for deriving CNN relevance maps may help to fill this gap as they allow the visualization of key input image features that drive the decision of the model. We investigated whether models with higher accuracy also rely more on discriminative brain regions predefined by prior knowledge.

**Methods:** We trained a CNN for the detection of AD in N=663 T1-weighted MRI scans of patients with dementia and amnestic mild cognitive impairment (MCI) and verified the accuracy of the models via cross-validation and in three independent samples including in total N=1655 cases. We evaluated the association of relevance scores and hippocampus volume to validate the clinical utility of this approach. To improve model comprehensibility, we implemented an interactive visualization of 3D CNN relevance maps, thereby allowing intuitive model inspection.

**Results:** Across the three independent datasets, group separation showed high accuracy for AD dementia versus controls (AUC≥0.91) and moderate accuracy for amnestic MCI versus controls (AUC≈0.74). Relevance maps indicated that hippocampal atrophy was considered as the most informative factor for AD detection, with additional contributions from atrophy in other cortical and subcortical regions. Relevance scores within the hippocampus were highly correlated with hippocampal volumes (Pearson's r ≈ –0.86, p<0.001).

**Conclusion:** The relevance maps highlighted atrophy in regions that we had hypothesized a priori. This strengthens the comprehensibility of the CNN models, which were trained in a purely data-driven manner based on the scans and diagnosis labels. The high hippocampus relevance scores as well as the high performance achieved in independent samples support the validity of the CNN models in the detection of AD-related MRI abnormalities. The presented data-driven and hypothesis-free CNN modeling approach might provide a useful tool to automatically derive discriminative features for complex diagnostic tasks where clear clinical criteria are still missing, for instance for the differential diagnosis between various types of dementia.

# Declarations

### Ethics approval and consent to participate

Data collecting within ADNI and AIBL was approved by participating institutions. See https://adni.loni.usc.edu and https://aibl.csiro.au for details. The DELCODE study was approved by participating institutions. See [1] for details. All study participants or their representatives provided written informed consent to participate in the respective studies and also agreed to sharing of their data. The retrospective analysis, study design and interactive visualization of relevance maps were approved by the internal review board of the Rostock University Medical Center, reference number A 2020-0182.

## Consent for publication

Not applicable.

## Availability of data and materials

Data used for training/evaluation of the models is available from the respective initiatives (ADNI: http://adni.loni.usc.edu/data-samples/access-data, AIBL: https://aibl.csiro.au, DELCODE: https://www.dzne.de/en/research/studies/clinical-studies/delcode).
The source code, a demo dataset, the trained CNN models, and all additional files required to run the interactive visualization are publicly available at GitHub:
https://github.com/martindyrba/DeepLearningInteractiveVis

## Competing interests

The authors declare that they have no competing interests.

## Funding

This study was supported by the Deutsche Forschungsgemeinschaft (DFG, German Research Foundation), project ID 454834942, funding code DY151/2-1.

## Authors' contributions

MD: conceptualization, methodology, data curation and processing, coding and software development, visualization, writing of the original draft. MH: coding and software development, visualization. ST: conceptualization, methodology, data collection and curation, writing, review and editing, supervision and clinical validation. All others: data acquisition, collection and curation, substantial intellectual contribution on study design and methodology, review and editing, clinical validation. All authors read and approved the final manuscript.

## Acknowledgements

The data samples were provided by the DELCODE study group of the Clinical Research Unit of the German Center for Neurodegenerative Diseases (DZNE). Details and participating sites can be found at www.dzne.de/en/research/studies/clinical-studies/delcode. The DELCODE study was supported by Max Delbrück Center for Molecular Medicine in the Helmholtz Association (MDC), Berlin; Center for Cognitive Neuroscience Berlin (CCNB) at Freie Universität Berlin; Bernstein Center for Computational Neuroscience (BCCN), Berlin; Core Facility MR-Research in Neurosciences, University Medical Center Goettingen; Institute for Clinical Radiology, Ludwig Maximilian University, Munich; Institute of Diagnostic and Interventional Radiology, Pediatric Radiology and Neuroradiology, Rostock University Medical Center; and Magnetic Resonance research center, University Hospital Tuebingen.
Data collection and sharing for this project was funded by the Alzheimer's Disease Neuroimaging Initiative (ADNI) (National Institutes of Health Grant U01 AG024904). ADNI is funded by the National Institute on Aging, the National Institute of Biomedical Imaging and Bioengineering, and through generous contributions from the following: AbbVie, Alzheimer's Association; Alzheimer's Drug Discovery Foundation; Araclon Biotech; BioClinica, Inc.; Biogen; Bristol-Myers Squibb Company; CereSpir, Inc.; Cogstate; Eisai Inc.; Elan Pharmaceuticals, Inc.; Eli Lilly and Company; EuroImmun; F. Hoffmann-La Roche Ltd and its affiliated company Genentech, Inc.; Fujirebio; GE Healthcare; IXICO Ltd.; Janssen Alzheimer Immunotherapy Research & Development, LLC.; Johnson & Johnson Pharmaceutical Research & Development LLC.; Lumosity; Lundbeck; Merck & Co., Inc.; Meso Scale Diagnostics, LLC.; NeuroRx Research; Neurotrack Technologies; Novartis Pharmaceuticals Corporation; Pfizer Inc.; Piramal Imaging; Servier; Takeda Pharmaceutical Company; and Transition Therapeutics. The Canadian Institutes of Health Research is providing funds to support ADNI clinical sites in Canada. Private sector contributions are facilitated by the Foundation for the National Institutes of Health (www.fnih.org). The grantee organization is the Northern California Institute for Research and Education, and the study is coordinated by the Alzheimer's Therapeutic Research Institute at the University of Southern California. ADNI data are disseminated by the Laboratory for Neuro Imaging at the University of Southern California.

# 1 Introduction

Alzheimer's disease (AD) is characterized by widespread neuronal degeneration, which manifests macroscopically as cortical atrophy that can be detected *in vivo* using structural magnetic resonance imaging (MRI) scans. Particularly at earlier stages of AD, atrophy patterns are relatively regionally specific, with volume loss in the medial temporal lobe and particularly the hippocampus. Therefore, hippocampus volume is currently the best-established MRI marker for diagnosing Alzheimer's disease at the dementia stage as well as at its prodromal stage amnestic mild cognitive impairment (MCI) [2, 3]. Automated detection of subtle brain changes in early stages of Alzheimer's disease could improve diagnostic confidence and early access to intervention [2, 4].

Convolutional neural networks (CNNs) provide a powerful method for image recognition. Various studies have evaluated the performance of CNNs for the detection of Alzheimer's disease in MR images with promising results regarding both separation of diagnostic groups and the prediction of conversion from MCI to manifest dementia. Despite the high accuracy levels achieved by CNN models, a major drawback is their algorithmic complexity, which renders them black-box systems. The poor intuitive comprehensibility of CNNs is one of the major obstacles which hinder the clinical application.

Novel methods for deriving relevance maps from CNN models [5, 6] may help to overcome the black-box problem. In general, relevance or saliency maps indicate the amount of information or contribution of a single input feature on the probability of a particular output class. Previous methodological approaches like gradient-weighted class activation mapping (Grad-CAM) [7], occlusion sensitivity analyses [8, 9], and local interpretable model-agnostic explanations (LIME) [10] had the limitation that deriving the relevance or saliency maps provided only group-average estimates, required long runtime [11] or provided only low spatial resolution [12, 13]. In contrast, more recent methods such as guided backpropagation [14] or layer-wise relevance propagation (LRP) [5, 6] use back-tracing of neural activation through the network paths to obtain high-resolution relevance maps.

Recently, three studies compared LRP with other CNN visualization methods for the detection of Alzheimer's disease in T1-weighted MRI scans [12, 13, 15]. The derived relevance maps showed the strongest contribution of medial and lateral temporal lobe atrophy, which matched the *a priori* expected brain regions of high diagnostic relevance [16, 17]. These preliminary findings provided the first evidence that CNN models and LRP visualization could yield reasonable relevance maps for individual people. We investigated whether this approach could be used as the basis for neuroradiological assistance systems to support the examination and diagnostic evaluation of MRI scans. Furthermore, we wanted to develop a data-driven and hypothesis-free CNN modeling approach that is capable of automatically deriving discriminative features and, therefore, might support complex diagnostic tasks where clear clinical criteria are still missing such as the differential diagnosis of various types of dementia.

In the current study, our aims were threefold: First, we trained robust CNN models that achieved a high diagnostic accuracy in three independent validation samples. Second, we developed a visualization software to interactively derive and inspect diagnostic relevance maps from CNN models for individual patients. Here, we expected high relevance to be shown in brain regions with strong disease-related atrophy, primarily in the medial temporal lobe. Third, we evaluated the validity of relevance maps in terms of correlation of hippocampus relevance scores and hippocampus volume, which is the best-established MRI marker for Alzheimer's disease [16, 17]. We expected a high consistency of both measures, which would strengthen the overall comprehensibility of the CNN models.

# 2 State of the art

## 2.1 Neural network models to detect Alzheimer's disease

An overview of neuroimaging studies which applied neural networks in the context of AD is provided in Table 1. We focused on the aspects whether the studies used independent validation samples to assess the generalizability of their models and whether they evaluated which image features contributed to the models' decision. Studies reported very high classification performances to differentiate AD dementia patients and cognitively healthy participants, typically with accuracies around 90 % (Table 1). For the separation of MCI and controls, accuracies were substantially lower ranging between 75 % and 85 %. However, there is a high variation of the accuracy levels depending on various factors such as i) differences in diagnostic criteria across samples, ii) included data types, iii) differences in image preprocessing procedures, and iv) differences between machine learning methods [18].

CNN performance estimation and model robustness are still open challenges. Wen and colleagues [18] actually showed only a minor effect of the particular CNN model parameterization or network layer configuration on the final accuracy, which means that the fully trained CNN models achieved almost identical performance. Different CNN approaches exist for MRI data [18] based on i) 2D convolutions for single slices, often reusing pre-trained models for general image detection, such as AlexNet [19] and VGG [20]; ii) so-called 2.5D approaches running 2D convolutions on each of the three slice orientations, which are then combined at higher layers of the network; and iii) 3D convolutions, which are at least theoretically superior in detecting texture and shape features in any direction of the 3D volume. Although final accuracy is almost comparable between all three approaches for detecting MCI and AD [18], the 3D models require substantially more parameters to be estimated during training. For instance, a single 2D convolutional kernel has 3×3=9 parameters whereas the 3D version requires 3×3×3=27 parameters. Here, relevance maps and related methods enable the assessment of learnt CNN models with respect to overfitting to clinically irrelevant brain regions and the detection of potential biases present in the training samples, which cannot be directly identified just from the model accuracy.

## 2.2 Approaches to assess model comprehensibility

In the literature, the most often applied methods to assess model comprehensibility and sensitivity were i) the visualization of model weights, ii) occlusion sensitivity analysis, and iii) more advanced CNN methods such as guided backpropagation or LRP (Table 1). Notably, studies using the approaches (i) and (ii) showed visualizations characterizing the whole sample or group averages. In contrast, studies applying (iii) also presented relevance maps for single participants [12, 15].

Böhle and colleagues [15] pioneered the application of LRP in neuroimaging and reported a high sensitivity of this method to actual regional atrophy. Eitel and colleagues [13] assessed the stability and reproducibility of CNN performance results and LRP relevance maps. After training ten individual models based on the same training dataset, they reported the highest consistency and lowest deviation of relevance maps for LRP and guided backpropagation among five different methods [13]. Recently, we compared various methods for relevance and saliency attribution [12]. Visually, all tested methods provided similar relevance maps except for Grad-CAM, which provided much lower spatial resolution, and, hence, lost a high amount of regional specificity. For the other methods, the main difference was the amount of "negative" relevance which indicates evidence against a particular diagnostic class. Notably, [13] and [15] did not include patients in the prodromal stage of MCI and [12] focused on a limited range of coronal slices covering the temporal lobe. All three studies did not validate their results in independent samples.

**Table 1** Overview of previous studies applying neural networks for the detection of AD and MCI.

| Study (chronologic order) | Data type | Sample | | | Algorithm | Performance | | | | Addressed model comprehensibility |
|---|---|---|---|---|---|---|---|---|---|---|
| | | AD | MCI c/nc | CN | | Groups | Accuracy | Balanced accuracy | AUC | |
| Suk et al. (2014) [21] | MRI GM and FDG-PET | 93 | 76/128 | 101 | RBM on class discriminative patches selected by statistical significance tests | AD/CN MCI/CN MCIc/MCInc | 95.4 % 85.7 % 74.6 % | 94.9 % 80.6 % 71.6 % | 0.988 0.881 0.747 | Visualization of selected features (image patches) and RBM model weights projected on MRI scan |
| Li et al. (2015) [22] | MRI and FDG-PET | 51 | 43/56 | 52 | RBM for feature learning, SVM for classification | AD/CN MCI/CN MCIc/MCInc | 91.4 % 77.4 % 57.4 % | | | No |
| Ortiz et al. (2016) [23] | MRI GM and FDG-PET | 70 | 39/64 | 68 | RBM for feature learning, SVM for classification | AD/CN MCIc/CN MCIc/MCInc | 90 % 83 % 78 % | | 0.95 0.95 0.82 | Visualization of SVM model weights projected on MRI scan |
| Aderghal et al. (2018) [24] | MRI and DTI | 188 | 339 | 228 | CNN for hippocampus region of interest only | AD/CN MCI/CN | 92.5 % 80.0 % | 92.5 % 82.9 % | | No |
| Liu et al. (2018a) [25] | FDG-PET | 93 | 146 | 100 | CNN and RNN | AD/CN MCI/CN | 91.2 % 78.9 % | | 0.953 0.839 | Visualization of most contributing brain areas obtained from occlusion sensitivity analysis |
| Liu et al. (2018b) [26] | MRI | 199 | – | 229 | CNN on landmarks selected by statistical significance tests | AD/CN MCIc/CN | 90.6 % | | 0.957 | Visualization of top 50 anatomical landmarks used as input for the CNN |
| Lin et al. (2018) [27] | MRI | 188 | 169/193 | 229 | CNN | AD/CN MCIc/MCInc | 88.8 % 79.9 % | | 0.861 | No |
| Böhle et al. (2019) [15] | MRI | 211 | – | 169 | CNN | AD/CN | 88.0 % | | | Visualization of LRP relevance and guided backpropagation maps, comparison of LRP relevance scores by group and brain region |
| Li et al. (2019) [28] | MRI | Training: 192 Test: 225 | 383 479 | 228 639 | CNN for hippocampus only | AD/CN MCIc/MCInc | 92.9 % | | 0.958 0.891 | Visualization of most contributing hippocampus areas obtained from CNN class activation mapping |
| Dyrba et al. (2020) [12] | MRI | 189 | 219 | 254 | CNN for coronal slices covering hippocampus | AD/CN MCI/CN | | | 0.93 0.75 | Visualization of LRP and other methods' relevance maps and comparison by diagnostic group |
| Lian et al. (2020) [29] | MRI | Training: 199 Test: 159 | 167/226 38/239 | 229 200 | CNN | AD/CN MCIc/MCInc | 90.3 % 80.9 % | | 0.951 0.781 | Visualization of most contributing image areas obtained from CNN class activation mapping |
| Qiu et al. (2020) [30] | MRI | Training: 188 Test$_1$: 62 Test$_2$: 29 Test$_3$: 209 | – – – – | 229 320 73 356 | FCN | AD/CN$_1$ AD/CN$_2$ AD/CN$_3$ | 87.0 % 76.6 % 81.8 % | | 0.870 0.892 0.881 | Visualization of most contributing brain areas obtained from occlusion sensitivity analysis |

| Wen et al. (2020) [18] | MRI | Training: 336<br>Test$_1$: 76<br>Test$_2$: 78 | 295/298<br>20/13<br>– | 330<br>429<br>76 | CNN | AD/CN$_1$<br>MCIc/MCInc$_1$<br>AD/CN$_2$ | 86 %<br>50 %<br>70 % | | No |
|---|---|---|---|---|---|---|---|---|---|
| Thibeau-Sutre et al. (2020) [9] | MRI | Training: 336<br>Test: 76 | –<br>– | 330<br>429 | CNN | AD/CN | 90 % | | Visualization of most contributing brain areas obtained from occlusion sensitivity analysis |
| Jo et al. (2020) [31] | Tau-PET | 66 | – | 66 | CNN | AD/CN | | 90.8 % | Visualization of LRP relevance maps, visualization of most contributing brain areas obtained from occlusion sensitivity analysis |

Empty cells in the performance columns indicate that the respective values were not reported.

AD: Alzheimer's dementia, MCI: mild cognitive impairment, MCIc: MCI converted to dementia, MCInc: non-converter/stable MCI, CN: cognitively normal controls, DTI: diffusion tensor imaging, FCN: fully connected network, RBM: restricted Boltzmann machine, RNN: recurrent neural network, CNN: convolutional neural network, MRI: T1-weighted magnetic resonance imaging, GM: gray matter volume, FDG-PET: glucose metabolism derived from fluorodeoxyglucose positron emission tomography.

# 3 Materials and Methods

## 3.1 Study samples

Data for **training** the CNN models were obtained from the Alzheimer's Disease Neuroimaging Initiative (ADNI) database (https://adni.loni.usc.edu). The ADNI was launched in 2003 by the National Institute on Aging, the National Institute of Biomedical Imaging and Bioengineering, the Food and Drug Administration, private pharmaceutical companies, and non-profit organizations, with the primary goal of testing whether neuroimaging, neuropsychological, and other biological measurements can be used as reliable in vivo markers of Alzheimer's disease pathogenesis. A complete description of ADNI, up-to-date information, and a summary of diagnostic criteria are available at https://www.adni-info.org. We selected a sample of N=663 participants from the ADNI-GO and ADNI-2 phases, based on the availability of concurrent T1-weighted MRI and amyloid AV45-PET scans. Notably, we used only one (i.e., the first) available scan from each ADNI participant in our analyses. The sample characteristics are shown in Table 2. We included 254 cognitively normal controls, 220 patients with (late) amnestic mild cognitive impairment (MCI), and 189 patients with Alzheimer's dementia (AD). Amyloid-beta status of the participants was determined by the UC Berkeley [32] based on the AV45-PET standardized uptake value ratio (SUVR) cutoff 1.11.

For **validation** of the diagnostic accuracy of the CNN models, we obtained MRI scans from three independent cohorts. The sample characteristics and demographic information are summarized in Table 2. The first dataset was compiled from N=575 participants of the recent ADNI-3 phase. The second dataset included MR images from N=606 participants of the Australian Imaging, Biomarker & Lifestyle Flagship Study of Ageing (AIBL) (https://aibl.csiro.au), provided via the ADNI system. A summary of the diagnostic criteria and additional information is available at https://aibl.csiro.au/about. For AIBL, we additionally obtained amyloid PET scans which were available for 564 participants (93%). The PET scans were processed using the Centiloid SPM pipeline and converted to Centiloid values as recommended for the different amyloid PET traces [33-35]. Amyloid-beta status of the participants was determined using the cutoff 24.1 CL [34]. As a third sample, we included data from N=474 participants of the German Center for Neurodegenerative Diseases (DZNE) multicenter observational study on Longitudinal Cognitive Impairment and Dementia (DELCODE) [1]. Comprehensive information on the diagnostic criteria and study design are provided in [1]. For the DELCODE sample, cerebrospinal fluid (CSF) biomarkers were available for a subsample of 227 participants (48%). Amyloid-beta status was determined using the Aβ42/Aβ40 ratio with a cutoff 0.09 [1].

**Table 2** Summary of sample characteristics.

| Sample | CN | MCI | AD |
|---|---|---|---|
| **ADNI-GO/2 (Training) N=663** | | | |
| Sample size (female) | 254 (130) | 220 (93) | 189 (80) |
| Age (SD) | 75.4 (6.6) | 74.1 (8.1) | 75.0 (8.0) |
| Education (SD) | 16.4 (2.7) | 16.2 (2.8) | 15.9 (2.7) |
| MMSE (SD) | 29.1 (1.2) | 27.6 (1.9) | 22.6 (3.2) |
| RAVLT Delayed recall (SD) | 7.6 (4.1) | 3.2 (3.7) | 0.8 (1.9) |
| WMS-LM Delayed recall (SD) | 13.9 (3.7) | 5.1 (3.8) | 1.5 (2.1) |
| Hippocampus volume (SD) mm³ | 6,235 (756) | 5,619 (963) | 4,834 (930) |
| Amyloid status (neg/pos) | 177/77 | 79/141 | 28/161 |
| MRI field strength (1.5T/3T) | 71/183 | 49/171 | 35/154 |
| **ADNI-3 (Validation) N=575** | | | |
| Sample size (female) | 326 (211) | 187 (85) | 62 (27) |
| Age (SD) | 70.0 (7.5) | 72.2 (7.5) | 74.8 (7.7) |
| Education (SD) | 16.6 (2.2) | 16.6 (2.5) | 16.5 (2.4) |
| MMSE (SD) | 29.1 (1.1) | 27.8 (2.0) | 23.1 (3.3) |
| RAVLT Delayed recall (SD) | 8.3 (4.4) | 4.7 (4.7) | 0.3 (0.9) |
| WMS-LM Delayed recall (SD) | 13.0 (3.5) | 7.2 (3.9) | 2.0 (2.8) |
| Hippocampus volume (SD) mm³ | 6,583 (649) | 6,112 (902) | 4,839 (978) |
| Amyloid status (neg/pos) | 75/39 | 19/27 | 3/17 |
| MRI field strength (1.5T/3T) | 0/326 | 0/187 | 0/62 |
| **AIBL (Validation) N=606** | | | |
| Sample size (female) | 448 (260) | 96 (46) | 62 (36) |
| Age (SD) | 72.4 (6.2) | 74.3 (6.9) | 73.2 (7.3) |
| MMSE (SD) | 28.7 (1.2) | 27.0 (2.2) | 21.2 (5.3) |
| WMS-LM Delayed recall (SD) | 11.2 (4.3) | 4.9 (4.0) | 1.0 (1.9) |
| Hippocampus volume (SD) mm³ | 6,362 (704) | 5,712 (1,028) | 4,940 (1,055) |
| Amyloid status (neg/pos) | 316/101 | 34/54 | 6/53 |
| MRI field strength (1.5T/3T) | 55/393 | 7/89 | 2/60 |
| **DELCODE (Validation) N=474** | | | |
| Sample size (female) | 215 (124) | 155 (72) | 104 (61) |
| Age (SD) | 69.5 (5.5) | 73.0 (5.7) | 75.2 (6.2) |
| Education (SD) | 14.7 (2.7) | 14.0 (3.1) | 12.9 (3.1) |
| MMSE (SD) | 29.5 (0.8) | 27.8 (2.0) | 23.1 (3.2) |
| WMS-LM Delayed recall (SD) | 14.3 (3.6) | 7.4 (5.2) | 1.8 (2.8) |
| Hippocampus volume (SD) mm³ | 6,543 (679) | 5,665 (950) | 4,610 (944) |
| Amyloid status (neg/pos) | 58/28 | 30/57 | 5/49 |
| MRI field strength (1.5T/3T) | 0/215 | 0/155 | 0/104 |

Numbers indicate mean and standard deviation (SD) if not indicated otherwise.
Years of education were not available for the AIBL dataset. RAVLT delayed recall scores were not available for the AIBL and DELCODE samples.
CN: cognitively normal controls, MCI: amnestic mild cognitive impairment, AD: Alzheimer's dementia, SD: standard deviation, MMSE: Mini Mental State Examination, RAVLT: Rey Auditory Verbal Learning Test, WMS-LM: Wechsler Memory Scale Logical Memory Test, MRI: magnetic resonance imaging.

## 3.2 Image preparation and processing

All MRI scans were preprocessed using the Computational Anatomy Toolbox (CAT12, v9.6/r7487) [36] for Statistical Parametric Mapping 12 (SPM12, v12.6/r1450, Wellcome Centre for Human Neuroimaging, London, UK). Images were segmented into grey and white matter, spatially normalized to the default CAT12 brain template in Montreal Neurological Institute (MNI) reference space using the DARTEL algorithm, resliced to an isotropic voxel size of 1.5 mm, and modulated to adjust for expansion and shrinkage of the tissue. Initially and after all processing steps, all scans were visually inspected to check for image quality. In all scans, effects of the covariates age, sex, total intracranial volume (TIV) and scanner magnetic field strength (FS) were reduced using linear regression. This step was performed, as these factors are known to affect the voxel intensities or regional brain volume [37, 38]. For each voxel $vx_{ij}$, linear models were fitted on the healthy controls:

$$vx_{ij} = \beta_{i0} + \beta_{i1}age_j + \beta_{i2}sex_j + \beta_{i3}TIV_j + \beta_{i4}FS_j + \varepsilon_{ij} \qquad (1)$$

with $i$ being the voxel index, $j$ being the healthy participant index, $\beta_i$ being the respective model coefficients (for each voxel), and $\varepsilon_i$ being the error term or residual. Subsequently, the predicted voxel intensities were subtracted from all participants' gray matter maps to obtain the residual images:

$$res_{ij} = vx_{ij} - (\beta_{i0} + \beta_{i1}age_j + \beta_{i2}sex_j + \beta_{i3}TIV_j + \beta_{i4}FS_j) \qquad (2)$$

Notably, we performed the estimation process (1) only for the healthy ADNI-GO/2 participants. Then, (2) was applied to all other participants and the validation samples. This method was applied as brain volume, specifically in the temporal lobe and hippocampus, is substantially decreasing/shrinking in old age independently of the disease process [37, 38] and we expected this approach to increase accuracy. As sensitivity analysis, we also repeated CNN training on the raw gray matter volume maps for comparison. Patients with MCI and AD were combined into one disease-positive group. On the one hand, this was done as we observed a low sensitivity of machine learning models for MCI when trained only on AD cases, due to the much larger and more heterogeneous patterns of atrophy in AD than in MCI, where atrophy is specifically present in medial temporal and parietal regions [39]. On the other hand, combining both groups substantially increased the training sample, which was required to reduce the overfitting of the CNN models.

### 3.3 CNN model structure and training

The CNN layer structure was adapted from [15] and [18], which was inspired by the prominent 2D image detection networks AlexNet [19] and VGG [20]. The model was implemented in Python 3.7 with Keras 2.2.4 and Tensorflow 1.15. The layout is shown in Figure 1. The residualized/raw 3D images with a resolution of 100×100×120 voxels were fed as input into the neural network and processed by three consecutive convolution blocks including 3D convolutions (5 filters of 3×3×3 kernel size) with rectified linear activation function (ReLU), maximum pooling (2×2×2 voxel patches), and batch normalization layers (Figure 1). Then, three dropout (10%) and fully connected layers with ReLU activation followed, each consisting of 64, 32, and 2 neurons, respectively. The weights of last two layers were regularized with the L2 norm penalty. The last layer had the softmax activation function that rescaled the class activation values to likelihood scores. The network required approximately 700,000 parameters to be estimated.

The whole CNN pipeline was evaluated by stratified tenfold cross-validation, partitioning the ADNI-GO/2 sample into approximately 600 training and 60 test images with almost equal distribution of CN, MCI, and AD cases. Additionally, data augmentation was used. All images included in the respective training subsamples were flipped along the coronal (L/R) axis and also translated by ±10 voxels in each direction (x/y/z), yielding fourteen times increased number of samples per epoch of approximately 8350 images. The CNN model was then trained with the ADAM optimizer, applying the categorical cross-entropy loss function, the learning rate of 0.0001, and a batch size of 20. As the training groups sizes were imbalanced, we set class weights of 1.31 for controls and 0.81 for MCI/AD in order to circumvent biased predictions. The weights were determined using the formula $0.5n/n_i$ as recommended in the TensorFlow tutorial [40]. To select the optimal models during training, we set the number of epochs to ten and saved the model state (epoch) which performed best on the test partition. On a Windows 10 computer with Intel Core i5-9600 hexa-core CPU, 64 GB working memory, and NVIDIA GeForce GTX 1650 CUDA GPU, training took approximately 35 minutes per fold and 12 hours in total. All ten models were saved to disk for further inspection and validation. As control analysis, we also repeated the whole procedure based on the raw image data (normalized gray matter volumes) instead of using the residuals as CNN input. Here, we set the number of epochs to 20 due to slower convergence of the models.

We also trained CNN models on the whole ADNI-GO/2 sample for further evaluation. Here, we fixed the number of epochs to 4 for the residualized data and 8 for the raw data. These values provided the highest average accuracy and lowest loss in the previous cross-validation.

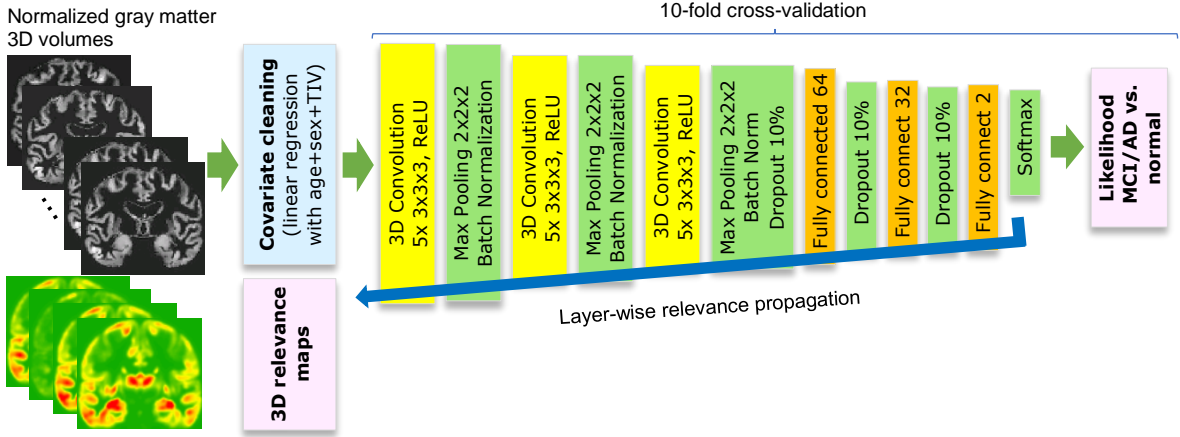

**Figure 1** Data flow chart and convolutional neural network structure.

### 3.4 Model evaluation

The balanced accuracy and area under the receiver operating characteristic curve (AUC) were calculated for the independent validation samples. We report first the numbers for the model trained on the whole ADNI-GO/2 dataset and second the average values for the models obtained via cross-validation.

As an internal validity benchmark, we compared CNN model performance and group separation using hippocampus volume, the best-established MRI marker for Alzheimer's disease. Automated extraction of hippocampus volume is already implemented in commercial radiology software to aid physicians in diagnosing dementia. We extracted total hippocampus volume from the modulated and normalized MRI scans using the Automated Anatomical Labeling (AAL) atlas [41]. The extracted volumes were corrected for the effects of age, sex, total intracranial volume, and magnetic field strength of the MRI scanner in the same way as described above for the CNN input (see the section 3.2). Here, a linear model was estimated based on the normal controls of the ADNI-GO/2 training sample, and then the parameters were applied to the measures of all other participants and validation samples to obtain the residuals. Subsequently, the residuals of the training sample were entered into a receiver operating characteristic analysis to obtain the AUC. The optimal threshold providing the highest accuracy was selected based on the Youden index. We obtained two thresholds. One for the separation of MCI and controls, which was the residual volume of –0.63 ml. That means participants with the deviation of individual hippocampus volume from the expected value (for that age, sex, total brain volume and magnetic field strength) below –0.63 ml were classified as MCI. The other threshold for AD dementia and controls was –0.95 ml. Additionally, we repeated the same cross-validation training/test splits as used for CNN training to compare the variability of the derived thresholds and performance measures.

### 3.5 CNN relevance map visualization

Relevance maps were derived from the CNN models using the LRP algorithm [5] implemented in the Python package iNNvestigate 1.0.9 [42]. LRP has previously been demonstrated to yield relevance maps with high spatial resolution and clinical plausibility [12, 15]. In this approach, the final network activation scores for a given input image are propagated back through the network layers. LRP applies a relevance conservation principle that means that the total amount of relevance per layer is kept constant during the back-tracing procedure to reduce numerical challenges that occur in other methods [5]. Several rules exist, which apply different weighting to positive (excitatory) and negative (inhibitory) connections such that network activation for and against a specific class can be considered differentially. Here, we applied the so-called α=1, β=0 rule that only considers positive relevance as proposed by [12, 15]. In this case, the relevance of a network neuron $R_j$ was calculated from all connected neurons $k$ in the subsequent network layer using the formula:

$$R_j = \sum_k \frac{a_j w_{jk}^+}{\sum_j (a_j w_{jk}^+)} R_k \tag{3}$$

with $a_j$ being the activation of neuron $j$, $w_{jk}^+$ being the positive weight of the connection between neurons $j$ and $k$, and $R_k$ being the relevance attributed to neuron $k$ [6]. As recent studies reported further improvements in LRP relevance attribution [43, 44], we applied the LRP α=1, β=0 composition rule that applies (3) to the convolutional layers, and the slightly extended ϵ rule [6] to the fully connected layers. In the ϵ rule, (3) is being extended by a small constant term added to the denominator, i.e. $\epsilon = 10^{-10}$ in our case, which is expected to reduce relevance when the activation of neuron k is weak or contradictory [6].

To facilitate model assessment and quick inspection of relevance maps, we implemented an interactive Python visualization application that is capable of immediate switching between CNN models and participants. More specifically, we used the Bokeh Visualization Library 2.2.3 (https://bokeh.org). Bokeh provides a webserver backend and web browser frontend to directly run Python code that dynamically generates interactive websites containing various graphical user interface components and plots. The Bokeh web browser JavaScript libraries handle the communication between the browser and server instance and translate website user interaction into Python function calls. In this way we implemented various visualization components to adjust plotting parameters and provide easy navigation for the 2D slice views obtained from the 3D MRI volume.

The application is structured following a model–view–controller paradigm. An overview of implemented functions is provided in Supplementary Figure 1. A sequence diagram illustrating function calls when selecting a new person is provided in Supplementary Figure 2. The source code and files required to run the interactive visualization are publicly available via https://github.com/martindyrba/DeepLearningInteractiveVis.

As core functionality we implemented the visualization in a classical 2D multi-slice window with axial, coronal and sagittal views, cross-hair, and sliders to adjust the relevance threshold as well as minimum cluster size threshold (see Figure 2 below). Here, a cluster refers to groups of adjacent voxels with high relevance above the selected relevance threshold. The cluster size is the number of voxels in this group, and can be controlled in order to reduce the visual noise caused by single voxels with high relevance. Additionally, we added visual guides to improve usability, including (a) a histogram providing the distribution of cluster sizes next to the cluster size threshold slider, (b) plots visualizing the amount of positive and negative relevance per slice next to the slice selection sliders, and (c) statistical information on the currently selected cluster. Further, assuming spatially normalized MRI data in MNI reference space, we added (d) atlas-based anatomical region lookup for the current cursor/cross-hair position and (e) the option to display the outline of the anatomical region to simplify visual comparison with the cluster location.

### 3.6 CNN model comprehensibility and validation

As quantitative metrics for assessing relevance map quality are still missing, we compared CNN relevance scores in the hippocampus with hippocampus volume. Here, we used the same AAL atlas hippocampus masks as for deriving hippocampus volume, and applied it on the relevance maps obtained from all ADNI-GO/2 participants for each model. The sum of relevance score of each voxel inside the mask was considered as hippocampus relevance. Hippocampus relevance and volume were compared using Pearson's correlation coefficient.

Additionally, we visually examined a large number of scans from each group to derive common relevance patterns and match them with the original MRI scans. Furthermore, we calculated mean relevance maps for each group. We also extracted the relevance for all lobes of the brain and subcortical structures to test the specificity of relevance distribution across the whole brain. These masks were defined based on the other regions included in the AAL atlas [41].

In an occlusion sensitivity analysis, we evaluated the influence of local atrophy on the prediction of the model and the derived relevance scores. Here, we slid a cube of 20 voxels = 30 mm edge size across the brain. Within the cube, we reduced the intensity of the voxel by 50 %, simulating gray matter atrophy in this area. We selected a normal control participant from the DELCODE dataset without visible CNN relevance, a prediction probability for AD/MCI of 20 %, and hippocampus volume residual of 0 ml, i.e. the hippocampus volume matched the reference volume expected for this person. For each position of the cube, we derived the probability of AD predicted by the model obtained from the whole ADNI-GO/2 sample. Additionally, we calculated the total amount of relevance in the scan.

# 4 Results

## 4.1 Group separation

The accuracy and AUC for diagnostic group separation are shown in Table 3. Additional performance measures are provided in Supplementary Table 1. The CNN reached a balanced accuracy between 75.5 % and 88.3 % across validation samples with an AUC between 0.828 and 0.978 for separating AD dementia and controls. For MCI vs. controls, the group separation was substantially lower with balanced accuracies between 63.1 % and 75.4 % and an AUC between 0.667 and 0.840. These values were only slightly better than the group separation performance of hippocampus volume (Table 3). The performance results for the raw gray matter volume data as input for the CNN are provided in Supplementary Table 2. In direct comparison to the CNN results for the residualized data, the balanced accuracies and AUC values did not show a clear difference (Table 3, Supplementary Table 2).

## 4.2 Model comprehensibility and relevance map visualization

The implemented web application frontend is displayed in Figure 2. The source code is available at https://github.com/martindyrba/DeepLearningInteractiveVis and the web application can be publicly accessed at https://explaination.net/demo. In the left column, the user can select a study participant and a specific model. Below, there are controls (sliders) to adjust the thresholds for displayed relevance score, cluster size, and overlay transparency. As we used the spatially normalized MRI images as CNN input, we can directly obtain the anatomical reference location label from the automated anatomical labeling (AAL) atlas [41] given the MNI coordinates at the specific cross-hair location, which is displayed in the light blue box. The green box displays statistics on the currently selected relevance cluster such as number of voxels and respective volume. In the middle part of Figure 2, the information used as covariates (age, sex, total intracranial volume, MRI field strength) and the CNN likelihood score for AD are depicted above the coronal, axial, and sagittal views of the 3D volume. We further added sliders and plots of cumulated relevance score per slices as visual guides to facilitate navigation to slices with high relevance. All user interactions are directly sent to the server, evaluated internally, and updated in the respective views and control components in real-time without major delay. For instance, adjusting the relevance threshold directly changes the displayed brain views, the shape of the red relevance summary plots, and the blue cluster size histogram. A sequence diagram of internal function calls when selecting a new participant is illustrated in Supplementary Figure 2.

**Table 3** Group separation performance for hippocampus volume and the convolutional neural network models.

| Sample | Hippocampus volume (residuals) | | 3D convolutional neural network | |
|---|---|---|---|---|
| | Balanced accuracy (mean ± SD) | AUC | Balanced accuracy (mean ± SD) | AUC (mean ± SD) |
| **ADNI-GO/2** | | | | |
| MCI vs. CN | *(70.0 % ± 6.8 %)* | *(0.773 ± 0.091)* | *(74.5 % ± 6.2 %)* | *(0.785 ± 0.078)* |
| AD vs. CN | *(84.4 % ± 3.6 %)* | *(0.945 ± 0.024)* | *(88.9 % ± 5.3 %)* | *(0.949 ± 0.029)* |
| MCI[+] vs. CN[-] | *(75.6 % ± 7.1 %)* | *(0.831 ± 0.080)* | *(86.7 % ± 10.3 %)* | *(0.925 ± 0.071)* |
| AD[+] vs. CN[-] | *(86.2 % ± 4.2 %)* | *(0.954 ± 0.025)* | *(94.9 % ± 3.8 %)* | *(0.985 ± 0.017)* |
| **ADNI-3** | | | | |
| MCI vs. CN | 62.8 % (63.1 % ± 1.4 %) | 0.683 | 63.1 % (63.6 % ± 1.5 %) | 0.684 (0.677 ± 0.020) |
| AD vs. CN | 83.4 % (83.4 % ± 0.4 %) | 0.917 | 84.4 % (81.7 % ± 2.9 %) | 0.913 (0.899 ± 0.013) |
| MCI[+] vs. CN[-] | 69.1 % (69.2 % ± 2.7 %) | 0.791 | 69.8 % (68.3 % ± 4.4 %) | 0.810 (0.742 ± 0.024) |
| AD[+] vs. CN[-] | 83.6 % (82.0 % ± 1.8 %) | 0.882 | 80.2 % (75.5 % ± 4.2 %) | 0.830 (0.828 ± 0.028) |
| **AIBL** | | | | |
| MCI vs. CN | 67.4 % (67.6 % ± 0.5 %) | 0.741 | 68.2 % (67.3 % ± 2.7 %) | 0.763 (0.749 ± 0.012) |
| AD vs. CN | 84.1 % (85.3 % ± 1.5 %) | 0.927 | 85.0 % (82.3 % ± 3.0 %) | 0.950 (0.926 ± 0.007) |
| MCI[+] vs. CN[-] | 78.5 % (78.8 % ± 0.9 %) | 0.874 | 75.4 % (73.6 % ± 3.1 %) | 0.828 (0.814 ± 0.022) |
| AD[+] vs. CN[-] | 87.2 % (89.1 % ± 2.4 %) | 0.976 | 88.3 % (85.3 % ± 3.3 %) | 0.978 (0.958 ± 0.011) |
| **DELCODE** | | | | |
| MCI vs. CN | 69.0 % (69.0 % ± 9.6 %) | 0.774 | 71.0 % (69.7 % ± 2.6 %) | 0.775 (0.772 ± 0.017) |
| AD vs. CN | 88.4 % (86.4 % ± 3.0 %) | 0.943 | 85.5 % (80.5 % ± 4.0 %) | 0.953 (0.938 ± 0.013) |
| MCI[+] vs. CN[-] | 77.4 % (77.8 % ± 0.7 %) | 0.867 | 72.2 % (74.9 % ± 3.5 %) | 0.840 (0.830 ± 0.017) |
| AD[+] vs. CN[-] | 88.2 % (87.6 % ± 1.8 %) | 0.954 | 83.3 % (82.2 % ± 4.0 %) | 0.968 (0.956 ± 0.012) |

Reported values are for the single model trained on the whole ADNI-GO/2 dataset. In parenthesis, the mean values and standard deviation for the ten models trained in the tenfold cross-validation procedure are provided to indicate the variability of the measures. Values for the ADNI-GO/2 sample (in italics) may be biased as the respective test subsamples were used to determine the optimal model during training. We still report them for better comparison of the model performance across samples.

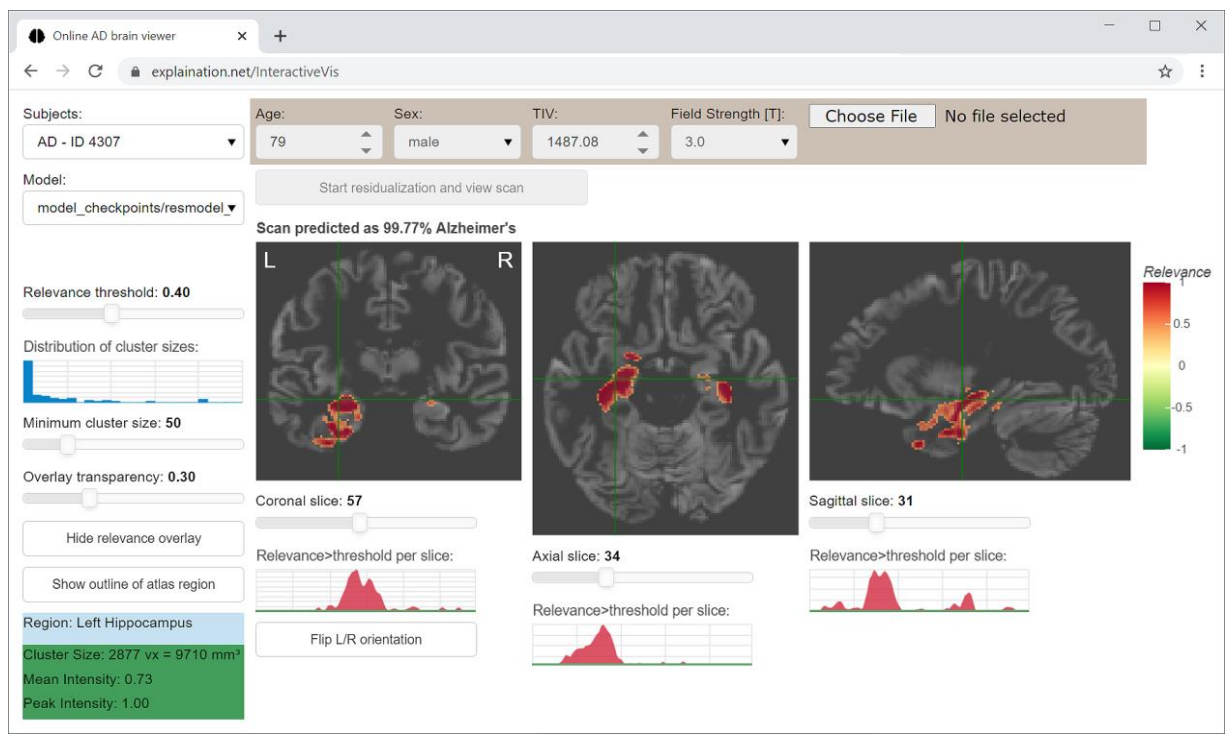

**Figure 2** Web application to interactively examine the neural network relevance maps for individual MRI scans.

Individual people's relevance maps are illustrated in Figure 3. The group mean relevance maps for the DELCODE validation sample are shown in Figure 4 and those for the ADNI-GO/2 training sample in Supplementary Figure 3. They are very similar to traditional statistical maps obtained from voxel-based morphometry, indicating the highest contribution of medial temporal brain regions, more specifically the hippocampus, amygdala, thalamus, middle temporal gyrus, and middle/posterior cingulate cortex. Also, they were highly consistent between samples (Supplementary Figure 3). The occlusion sensitivity analysis also showed identical brain regions' atrophy to contribute to the model's decision (Figure 5). Interestingly, the occlusion relevance maps showed a ring structure around the most contributing brain areas, indicating that relevance was highest when the occluded area just touched the salient regions, leading to a thinning-like shape of the gray matter.

The correlation of individual DELCODE participants' hippocampus relevance score and hippocampus volume for the model trained on the whole ADNI-GO/2 dataset is displayed in Figure 6. For this model, the correlation was r = –0.87 for bilateral hippocampus volume (p<0.001). Across all ten models obtained using cross-validation, the median correlation of total hippocampus relevance and volume was r = –0.84 with a range of –0.88 and –0.44 (all with p<0.001). Cross-validation models with higher correlation between hippocampus relevance and volume showed a tendency for better AUC values for MCI vs. controls (r = 0.61, p = 0.059). To test whether hippocampus volume and relevance measures were specific to the hippocampus, we also compared the correlation between hippocampus volume and other regions' and whole-brain relevance. Here the correlations were lower, with r = –0.62 (p<0.001) between hippocampus volume and whole-brain relevance. More detailed results are provided as a correlation matrix in Supplementary Figure 4.

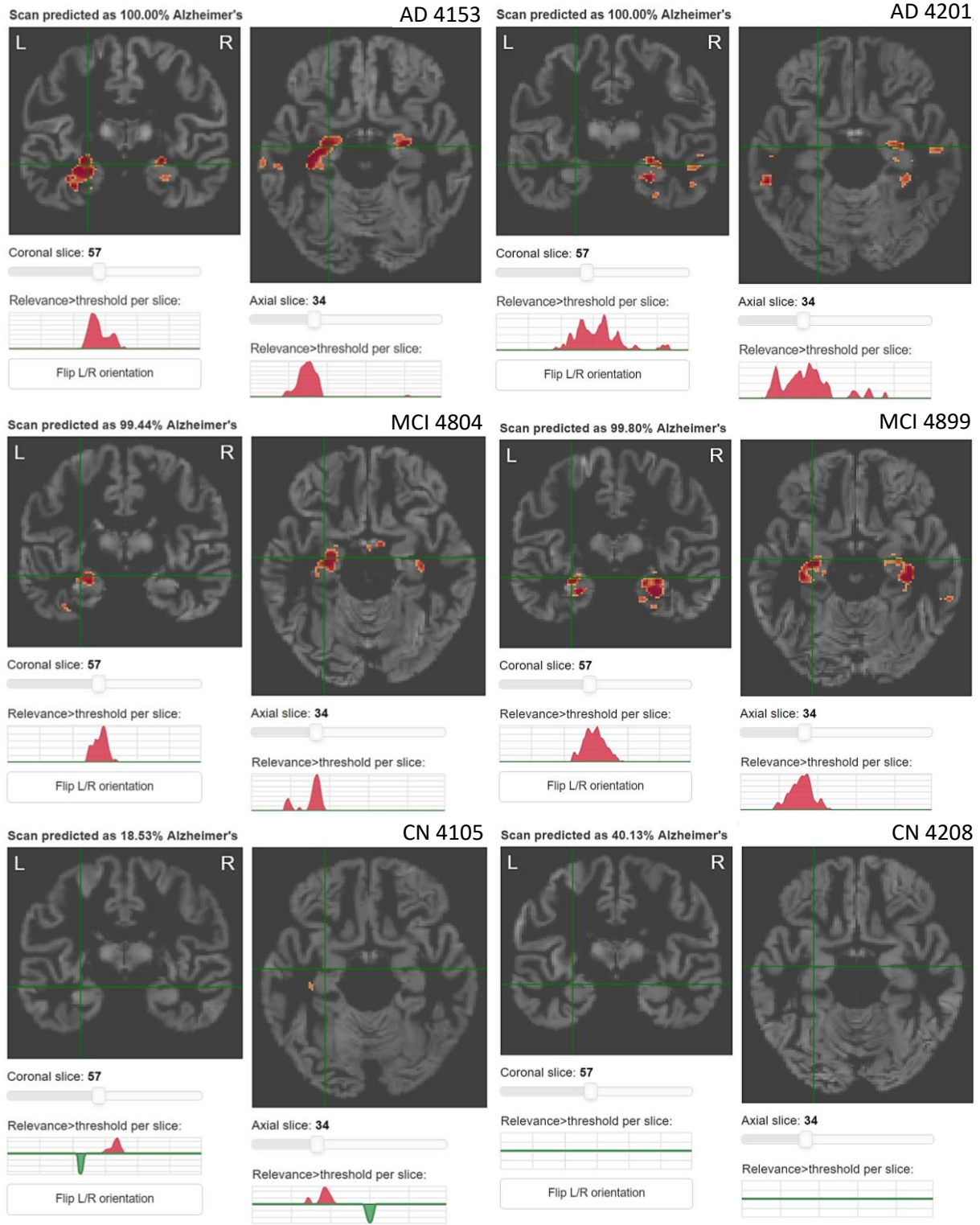

**Figure 3** Example relevance maps obtained for different people. Top row: Alzheimer's dementia patients, middle row: patients with mild cognitive impairment, bottom row: cognitively normal controls.

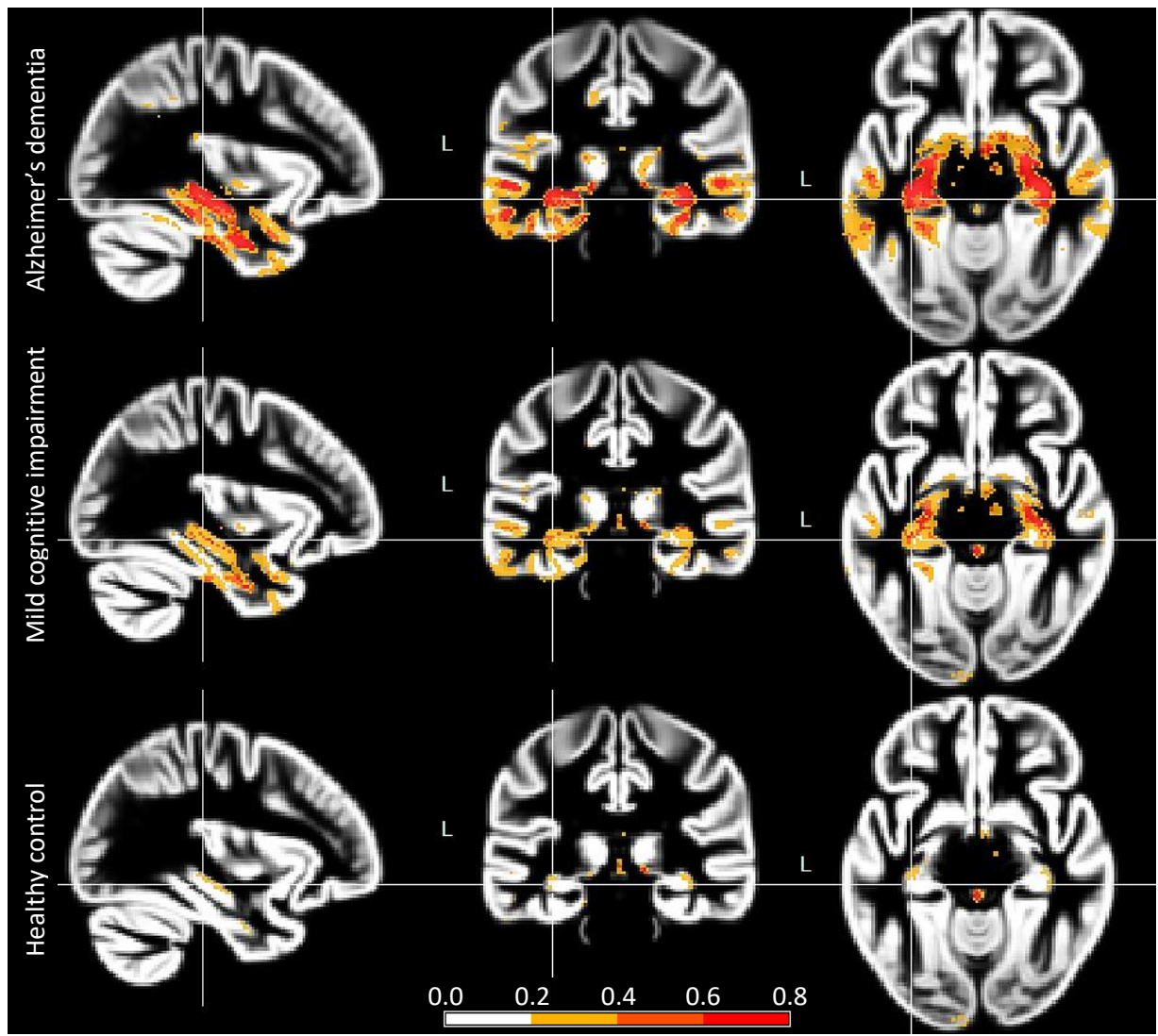

**Figure 4** Mean relevance maps for Alzheimer's dementia patients (top row), patients with mild cognitive impairment (middle row), and healthy controls (bottom row) for the DELCODE validation sample. Relevance maps thresholded at 0.2 for better comparison.

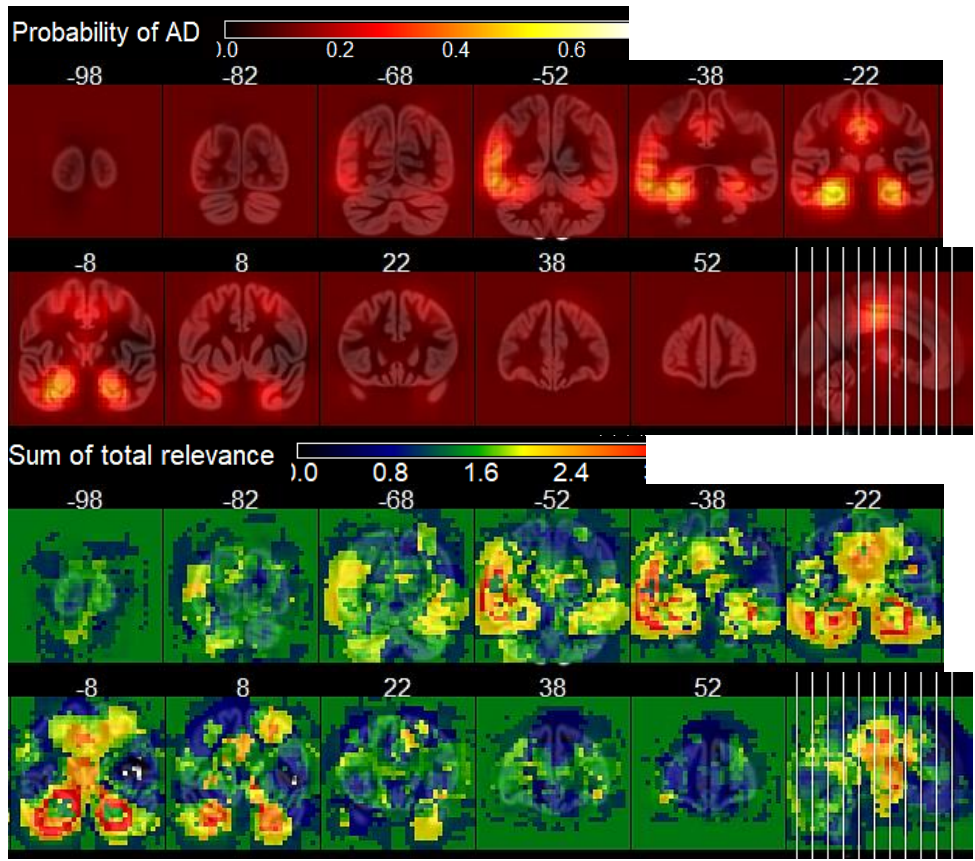

**Figure 5** Results from the occlusion sensitivity analysis. A gray matter volume loss of 50 % was simulated in a cube of 30 mm edge length. Each voxel encodes the derived values when centering the cube at that position. Top: Probability of AD for the areas with simulated atrophy. Bottom: Total sum of image relevance depending on simulated atrophy. Numbers indicate the y-axis slice coordinates in MNI reference space.

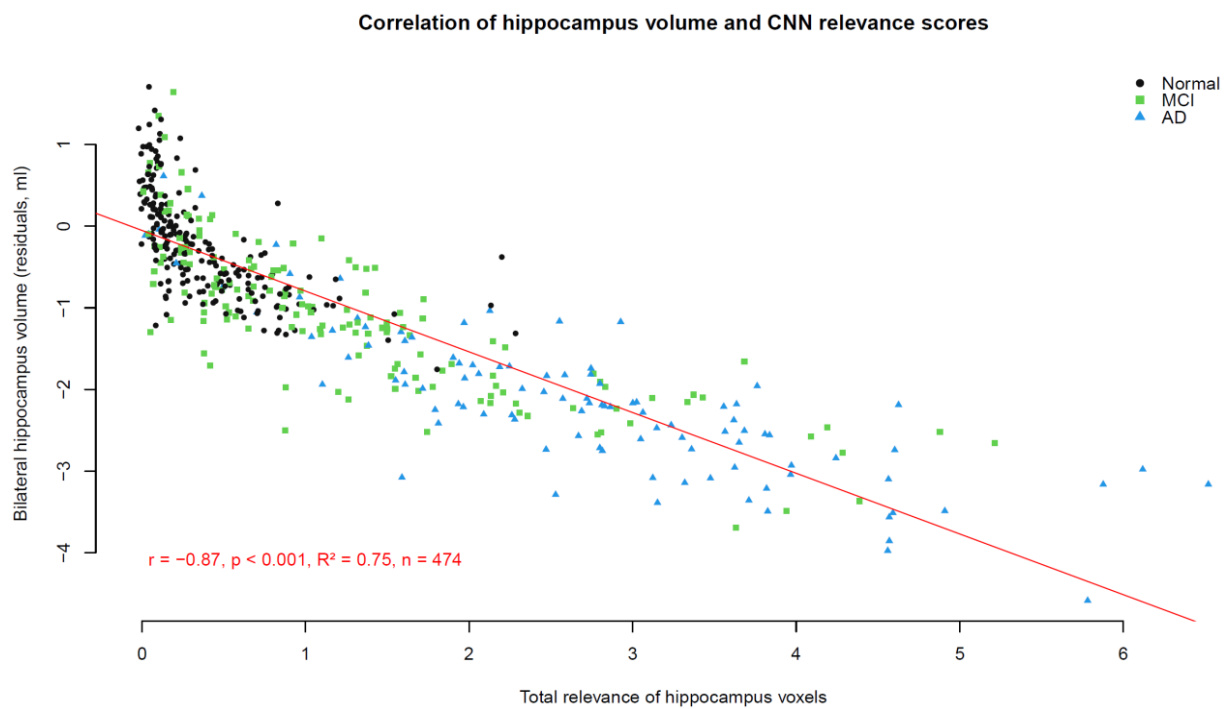

**Figure 6** Scatter plot and correlation of bilateral hippocampus volume and neural network relevance scores for the hippocampus region for the DELCODE sample (r = −0.87, p < 0.001).

# 5 Discussion
## 5.1 Neural network comprehensibility

We have presented a CNN framework and interactive visualization application for obtaining class-specific relevance maps for disease detection in MRI scans, yielding human-interpretable and clinically plausible visualizations of key features for image discrimination. To date, most CNN studies focus on model development and optimization, which are undoubtedly important tasks and there are still several challenges to tackle. However, as black-box models, it is typically not feasible to judge, why a CNN fails or which image features drive a particular decision of the network. This gap might be closed with the use of novel visualization algorithms such as LRP [5] and deep Taylor decomposition [6]. In our application, LRP relevance maps provided a useful tool for model inspection to reveal the brain regions which contributed most to the decision process encoded by the neural network models.

Currently, there is no ground truth information for relevance maps, and there are no appropriate methods available to quantify relevance map quality. Samek and colleagues [45] proposed the information-theoretic measures relevance map entropy and complexity, which mainly characterize the scatter or smoothness of images. Furthermore, adapted from classical neural network sensitivity analysis, they assessed the robustness of relevance maps using perturbation testing where small image patches were replaced by random noise, which was also applied in [46]. Already for 2D data, this method is computationally very expensive and only practical for a limited number of input images. Instead of adding random noise, we simulated gray matter atrophy by lowering the image intensities by 50 % in a cube-shaped area. As visible from Figure 5, the brain areas contributing to the model's AD probability nicely matched the areas shown in the mean relevance maps (Figure 4). Notably, the ring-shaped increase in relevance around the salient regions (Figure 5, bottom) indicates that the model is sensitive to intensity jumps occurring when the occlusion cube touches the borderline of those regions. Most probably, this means that the model was more sensitive to thinning patterns of gray matter than to equally-distributed volume reduction. However, our findings have to be seen as preliminary, as we only assessed this analysis in one normal control participant due to the computational effort, and therefore it requires more extensive research in future studies.

Based on the extensive knowledge about the effect of Alzheimer's disease on brain volume as presented in T1-weighted MRI scans [16, 17], we selected a direct quantitative comparison of relevance maps with hippocampus volume as validation method. Here, we obtained very high correlations between hippocampus relevance scores and volume (median correlation $r = -0.84$), underlining the clinical plausibility of learnt patterns to differentiate AD and MCI patients from controls. In addition, visual inspection of relevance maps also revealed several other clusters with gray matter atrophy in the individual participants' images that contributed to the decision of the CNN (Figure 2 and Figure 3). Böhle and colleagues [15] proposed an atlas-based aggregation of CNN relevance maps to be used as "disease fingerprints" and to enable a quick comparison between patients and controls, a concept that has also been proposed previously for differential diagnosis of dementia based on heterogeneous clinical data and other machine learning models [47, 48].

Notably, the CNN models presented here were solely based on the combinations of input images with their corresponding diagnostic labels to determine which brain features were diagnostically relevant. Traditionally, extensive clinical experience is required to define relevant features (e.g., hippocampus volume) that discriminate between a clinical population (here: AD, MCI) and a healthy control group. Also, typically, only few predetermined parameters are used (e.g., hippocampus volume or medial temporal lobe atrophy score [16, 17]). Our results demonstrate that the combination of CNN and relevance map approaches constitutes a promising tool for improving the utility of CNN in the classification of MRIs of patients with suspected AD in a clinical context. By referring back to the relevance maps, trained clinicians will be enabled to compare classification results to comprehensible features visible in the relevance images and thereby more readily interpret the classification results in clinically ambiguous situations. Perspectives, the relevance map approach might also provide a helpful tool to reveal features for more complex diagnostic challenges such as differential diagnosis between various types of dementia, for instance the differentiation between AD, frontotemporal dementia, and dementia with Lewy bodies.

### 5.2 CNN performance

As expected, CNN-based classification reached an excellent AUC ≥ 0.91 for the group separation of AD compared to controls but a substantially lower accuracy for group separation between MCI and controls (AUC ≈ 0.74, Table 3). When restricting the classification to amyloid-positive MCI versus amyloid-negative controls, group separation improved to AUC = 0.84 in DELCODE, highlighting the heterogeneity of MCI as a diagnostic entity and the importance of biomarker stratification [2, 3]. In summary, these numbers are also reflected by the recent CNN literature as shown in Table 1. Notably, [18] reported several limitations and issues in the performance evaluation of some other CNN papers, such that it is not easy to finally conclude on the group separation capabilities of the CNN models in realistic settings. To overcome such challenges, we validated the models on three large independent cohorts (Table 3), providing strong evidence for their generalizability and for the robustness of our CNN approach.

To put the CNN model performance into perspective, we compared the accuracy of the CNN models with the accuracy achieved by assessing hippocampus volume, the key clinical MRI marker for neurodegeneration in Alzheimer's disease [2, 3]. Interestingly, there were only minor differences in the achieved AUC values across all samples (Table 3). The MCI group of the ADNI-3 sample, which yielded the worst group separation of all samples (AUC = 0.68), was actually the group with the largest average hippocampus volumes and, therefore, the lowest group difference compared to the controls (Table 2). Obviously, our results here indicate a limited value of using CNN models instead of traditional volumetric markers for the detection of Alzheimer's dementia and mild cognitive impairment. Previous MRI CNN papers have not reported the baseline accuracy reached by hippocampus volume for comparison. However, as noted above, CNNs might provide a useful tool to automatically derive discriminative features for complex diagnostic tasks where clear clinical criteria are still missing, for instance for the differential diagnosis between various types of dementia.

### 5.3 Limitations

As already mentioned above, visual inspection of relevance maps also revealed several other regions with gray matter atrophy in the individual participants' images that contributed to the decision of the CNN. These additional regions were not further assessed, as a priori knowledge regarding their diagnostic value is still under debate in the scientific community [2, 3]. Also, we did not perform a three-way classification between AD dementia, MCI and CN due to the limited availability of cases for training. Additionally, MCI itself is a heterogeneous diagnostic entity [2, 3]. Here, all the studies involved in our analysis tried to increase the likelihood of underlying Alzheimer's pathology by focusing on MCI patients with memory impairment. But markers of amyloid-beta pathology were only available for a subset of participants such that we could not stratify by amyloid status for the training of the CNN models. However, we optionally applied this stratification for the validation of the CNN performances to improve the diagnostic confidence.

### 5.4 Future prospects

Several studies focused on CNN models for the integration of multimodal imaging data, e.g. MRI and fluorodeoxyglucose (FDG)-PET [21-23], or heterogeneous clinical data [49]. Here, it will be beneficial, to directly include the variables we used as covariates (such as age and sex) as input to the CNN model rather than performing the variance reduction directly on the input data before applying the model.

In this context, relevance mapping visualization approaches need to be developed that allow for a direct comparison of the relevance magnitude for images and clinical variables simultaneously. Another aspect is the automated generation of textual descriptions and diagnostic explanations from images [50-52]. Given the recent technical progress, we suggest that the approach is now ready for interdisciplinary exchange to assess how clinicians can benefit from CNN assistance in their diagnostic workup, and which requirements must be met to increase clinical utility. Beyond the technical challenges, regulatory and ethical aspects and caveats must be carefully considered when introducing CNN as part of clinical decision support systems and medical software—and the discussion of these issues has just recently begun [53, 54].

## 5.5 Conclusion

We presented a framework for obtaining diagnostic relevance maps from CNN models to improve model comprehensibility. These relevance maps have revealed reproducible and clinically plausible atrophy patterns in AD and MCI patients, with a high correlation with the well-established MRI marker of hippocampus volume. The implemented web application allows a quick and versatile inspection of brain regions with a high relevance score in individuals. With the increased comprehensibility of CNNs provided by the relevance maps, the data-driven and hypothesis-free CNN modeling approach might provide a useful tool to aid differential diagnosis of dementia and other neurodegenerative diseases, where fine-grained knowledge on discriminating brain alterations is still missing.

# List of abbreviations

AAL - automated anatomical labeling
AD – Alzheimer's disease
ADNI – Alzheimer's Disease Neuroimaging Initiative
AIBL - Australian Imaging, Biomarker & Lifestyle Flagship Study of Ageing
AUC - area under the receiver operating characteristic curve
CAM - class activation mapping
CI - confidence interval
CN - cognitively normal participants
CNN - convolutional neural network
CSF - cerebrospinal fluid
DELCODE - DZNE multicenter observational study on Longitudinal Cognitive Impairment and Dementia
DZNE - Deutsches Zentrum für Neurodegenerative Erkrankungen (German Center for Neurodegenerative Diseases)
FDG - fluorodeoxyglucose
GM - gray matter
LRP - layer-wise relevance propagation
MCI - mild cognitive impairment
MNI - Montreal Neurological Institute
MRI - magnetic resonance imaging
PET - positron emission tomography
ReLU - rectified linear activation function
SD - standard deviation
SUVR - standardized uptake value ratio
TIV - total intracranial volume

# Supplementary material

**Supplementary Table 1** Group separation performance for hippocampus volume and the convolutional neural network models for residualized data (extended).

| Sample | Hippocampus volume (residuals) | | 3D convolutional neural network | | | | | | |
|---|---|---|---|---|---|---|---|---|---|
| | Balanced accuracy (mean ± SD) | AUC | Balanced accuracy (mean ± SD) | Sensitivity (mean ± SD) | Specificity (mean ± SD) | F1-score (mean ± SD) | Positive predictive value (mean ± SD) | Negative predictive value (mean ± SD) | AUC (mean ± SD) |
| **ADNI-GO/2** | | | | | | | | | |
| MCI vs. CN | (70.0 % ± 6.8 %) | (0.773 ± 0.091) | (74.5 % ± 6.2 %) | (0.655 ± 0.108) | (0.836 ± 0.081) | (0.707 ± 0.080) | (0.781 ± 0.095) | (0.741 ± 0.063) | (0.785 ± 0.078) |
| AD vs. CN | (84.4 % ± 3.6 %) | (0.945 ± 0.024) | (88.9 % ± 5.3 %) | (0.942 ± 0.053) | (0.836 ± 0.081) | (0.872 ± 0.061) | (0.815 ± 0.084) | (0.952 ± 0.043) | (0.949 ± 0.029) |
| MCI[+] vs. CN[-] | (75.6 % ± 7.1 %) | (0.831 ± 0.080) | (86.7 % ± 10.3 %) | (0.790 ± 0.173) | (0.943 ± 0.042) | (0.843 ± 0.127) | (0.916 ± 0.067) | (0.850 ± 0.116) | (0.925 ± 0.071) |
| AD[+] vs. CN[-] | (86.2 % ± 4.2 %) | (0.954 ± 0.025) | (94.9 % ± 3.8 %) | (0.956 ± 0.038) | (0.943 ± 0.042) | (0.941 ± 0.043) | (0.927 ± 0.051) | (0.966 ± 0.029) | (0.985 ± 0.017) |
| **ADNI-3** | | | | | | | | | |
| MCI vs. CN | 62.8 % (63.1 % ± 1.4 %) | 0.683 | 63.1 % (63.6 % ± 1.5 %) | 0.421 (0.496 ± 0.082) | 0.850 (0.775 ± 0.080) | 0.492 (0.523 ± 0.031) | 0.611 (0.570 ± 0.060) | 0.716 (0.730 ± 0.015) | 0.684 (0.677 ± 0.020) |
| AD vs. CN | 83.4 % (83.4 % ± 0.4 %) | 0.917 | 84.4 % (81.7 % ± 2.9 %) | 0.839 (0.858 ± 0.036) | 0.850 (0.775 ± 0.080) | 0.638 (0.573 ± 0.068) | 0.515 (0.438 ± 0.088) | 0.965 (0.967 ± 0.006) | 0.913 (0.899 ± 0.013) |
| MCI[+] vs. CN[-] | 69.1 % (69.2 % ± 2.7 %) | 0.791 | 69.8 % (68.3 % ± 4.4 %) | 0.556 (0.615 ± 0.124) | 0.840 (0.752 ± 0.086) | 0.556 (0.523 ± 0.031) | 0.556 (0.479 ± 0.057) | 0.840 (0.847 ± 0.031) | 0.810 (0.742 ± 0.024) |
| AD[+] vs. CN[-] | 83.6 % (82.0 % ± 1.8 %) | 0.882 | 80.2 % (75.5 % ± 4.2 %) | 0.765 (0.759 ± 0.043) | 0.840 (0.752 ± 0.086) | 0.619 (0.573 ± 0.068) | 0.520 (0.424 ± 0.080) | 0.940 (0.932 ± 0.012) | 0.830 (0.828 ± 0.028) |
| **AIBL** | | | | | | | | | |
| MCI vs. CN | 67.4 % (67.6 % ± 0.5 %) | 0.741 | 68.2 % (67.3 % ± 2.7 %) | 0.552 (0.596 ± 0.111) | 0.812 (0.749 ± 0.086) | 0.455 (0.437 ± 0.020) | 0.387 (0.351 ± 0.057) | 0.894 (0.898 ± 0.016) | 0.763 (0.749 ± 0.012) |
| AD vs. CN | 84.1 % (85.3 % ± 1.5 %) | 0.927 | 85.0 % (82.3 % ± 3.0 %) | 0.887 (0.897 ± 0.051) | 0.812 (0.749 ± 0.086) | 0.547 (0.523 ± 0.055) | 0.396 (0.350± 0.090) | 0.981 (0.982 ± 0.007) | 0.950 (0.926 ± 0.007) |
| MCI[+] vs. CN[-] | 78.5 % (78.8 % ± 0.9 %) | 0.874 | 75.4 % (73.6 % ± 3.1 %) | 0.685 (0.713 ± 0.095) | 0.823 (0.759 ± 0.089) | 0.503 (0.464 ± 0.051) | 0.398 (0.356 ± 0.082) | 0.939 (0.940 ± 0.013) | 0.828 (0.814 ± 0.022) |
| AD[+] vs. CN[-] | 87.2 % (89.1 % ± 2.4 %) | 0.976 | 88.3 % (85.3 % ± 3.3 %) | 0.943 (0.947 ± 0.048) | 0.823 (0.759 ± 0.089) | 0.629 (0.573 ± 0.085) | 0.472 (0.420 ± 0.104) | 0.989 (0.989 ± 0.008) | 0.978 (0.958 ± 0.011) |
| **DELCODE** | | | | | | | | | |
| MCI vs. CN | 69.0 % (69.0 % ± 9.6 %) | 0.774 | 71.0 % (69.7 % ± 2.6 %) | 0.652 (0.724 ± 0.048) | 0.767 (0.670 ± 0.084) | 0.660 (0.664 ± 0.017) | 0.669 (0.618 ± 0.051) | 0.753 (0.771 ± 0.009) | 0.775 (0.772 ± 0.017) |
| AD vs. CN | 88.4 % (86.4 % ± 3.0 %) | 0.943 | 85.5 % (80.5 % ± 4.0 %) | 0.942 (0.939 ± 0.017) | 0.767 (0.670 ± 0.084) | 0.778 (0.719 ± 0.046) | 0.662 (0.585 ± 0.062) | 0.965 (0.958 ± 0.011) | 0.953 (0.938 ± 0.013) |
| MCI[+] vs. CN[-] | 77.4 % (77.8 % ± 0.7 %) | 0.867 | 72.2 % (74.9 % ± 3.5 %) | 0.737 (0.809 ± 0.046) | 0.707 (0.690 ± 0.085) | 0.724 (0.762 ± 0.027) | 0.712 (0.723 ± 0.049) | 0.732 (0.787 ± 0.031) | 0.840 (0.830 ± 0.017) |
| AD[+] vs. CN[-] | 88.2 % (87.6 % ± 1.8 %) | 0.954 | 83.3 % (82.2 % ± 4.0 %) | 0.959 (0.955 ± 0.021) | 0.707 (0.690 ± 0.085) | 0.832 (0.824 ± 0.033) | 0.734 (0.726 ± 0.053) | 0.953 (0.949 ± 0.023) | 0.968 (0.956 ± 0.012) |

Reported values are for the single model trained on the whole ADNI-GO/2 dataset. In parenthesis, the mean values and standard deviation for the ten models trained in the tenfold cross-validation procedure are provided to indicate the variability of the measures. Values for the ADNI-GO/2 sample (in italics) may be biased as the respective test subsamples were used to determine the optimal model during training. We still report them for better comparison of the model performance across samples.

**Supplementary Table 2** Group separation performance for hippocampus volume and the convolutional neural network models for raw input data.

| Sample | 3D convolutional neural network | |
|---|---|---|
| | Balanced accuracy (mean ± SD) | AUC (mean ± SD) |
| **ADNI-GO/2** | | |
| MCI vs. CN | *(71.1 % ± 5.7 %)* | *(0.731 ± 0.070)* |
| AD vs. CN | *(84.6 % ± 6.5 %)* | *(0.921 ± 0.024)* |
| MCI$^+$ vs. CN$^-$ | *(80.7 % ± 7.9 %)* | *(0.881 ± 0.069)* |
| AD$^+$ vs. CN$^-$ | *(92.4 % ± 3.9 %)* | *(0.974 ± 0.015)* |
| | | |
| **ADNI-3** | | |
| MCI vs. CN | 62.6 % (60.7 % ± 2.6 %) | 0.629 (0.626 ± 0.017) |
| AD vs. CN | 86.1 % (82.1 % ± 5.8 %) | 0.919 (0.907 ± 0.028) |
| MCI$^+$ vs. CN$^-$ | 71.8 % (70.6 % ± 4.9 %) | 0.769 (0.745 ± 0.021) |
| AD$^+$ vs. CN$^-$ | 82.2 % (78.8 % ± 5.2 %) | 0.873 (0.877 ± 0.026) |
| | | |
| **AIBL** | | |
| MCI vs. CN | 69.1 % (64.8 % ± 3.2 %) | 0.735 (0.713 ± 0.016) |
| AD vs. CN | 83.7 % (80.2 % ± 6.3 %) | 0.922 (0.924 ± 0.006) |
| MCI$^+$ vs. CN$^-$ | 78.0 % (73.3 % ± 4.5 %) | 0.837 (0.817 ± 0.025) |
| AD$^+$ vs. CN$^-$ | 86.3 % (83.7 % ± 6.8 %) | 0.959 (0.959 ± 0.007) |
| | | |
| **DELCODE** | | |
| MCI vs. CN | 69.8 % (69.0 % ± 2.4 %) | 0.779 (0.761 ± 0.017) |
| AD vs. CN | 89.8 % (83.5 % ± 6.0 %) | 0.947 (0.937 ± 0.023) |
| MCI$^+$ vs. CN$^-$ | 72.5 % (72.5 % ± 5.9 %) | 0.853 (0.814 ± 0.049) |
| AD$^+$ vs. CN$^-$ | 92.5 % (86.0 % ± 7.1 %) | 0.982 (0.967 ± 0.028) |

Reported values are the respective measures for the single model trained on the whole ADNI-GO/2 dataset. In parenthesis, the mean values and standard deviation for the ten models trained in the tenfold cross-validation procedure are provided to indicate the variability of the measures. Values for the ADNI-GO/2 sample (in italics) may be biased as the respective test subsamples were used to determine the optimal model during training. We still report them for better comparison of the model performance across samples.

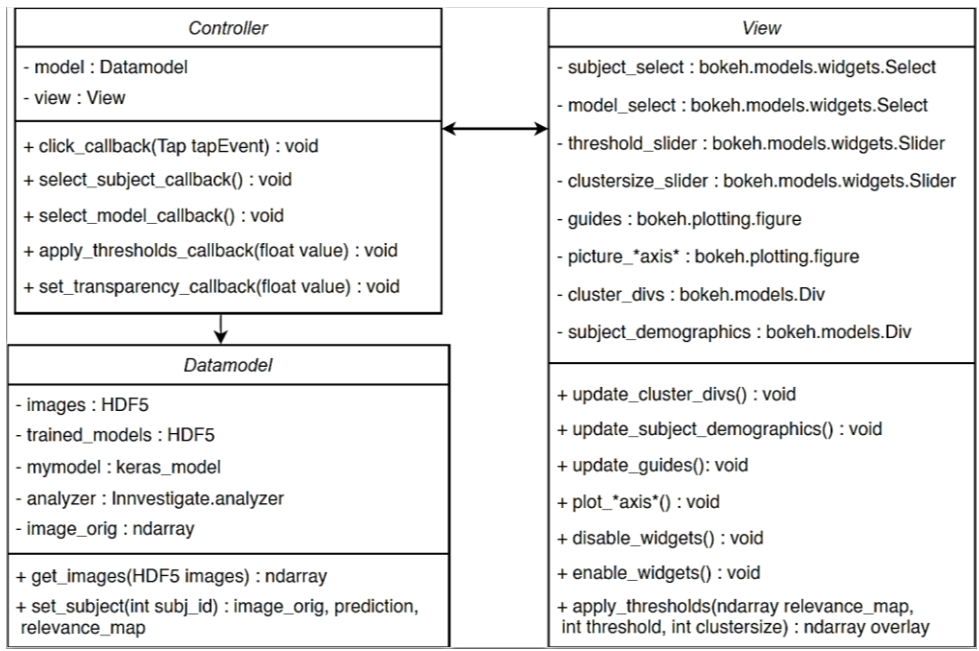

**Supplementary Figure 1**    UML diagram of the interactive visualization application.

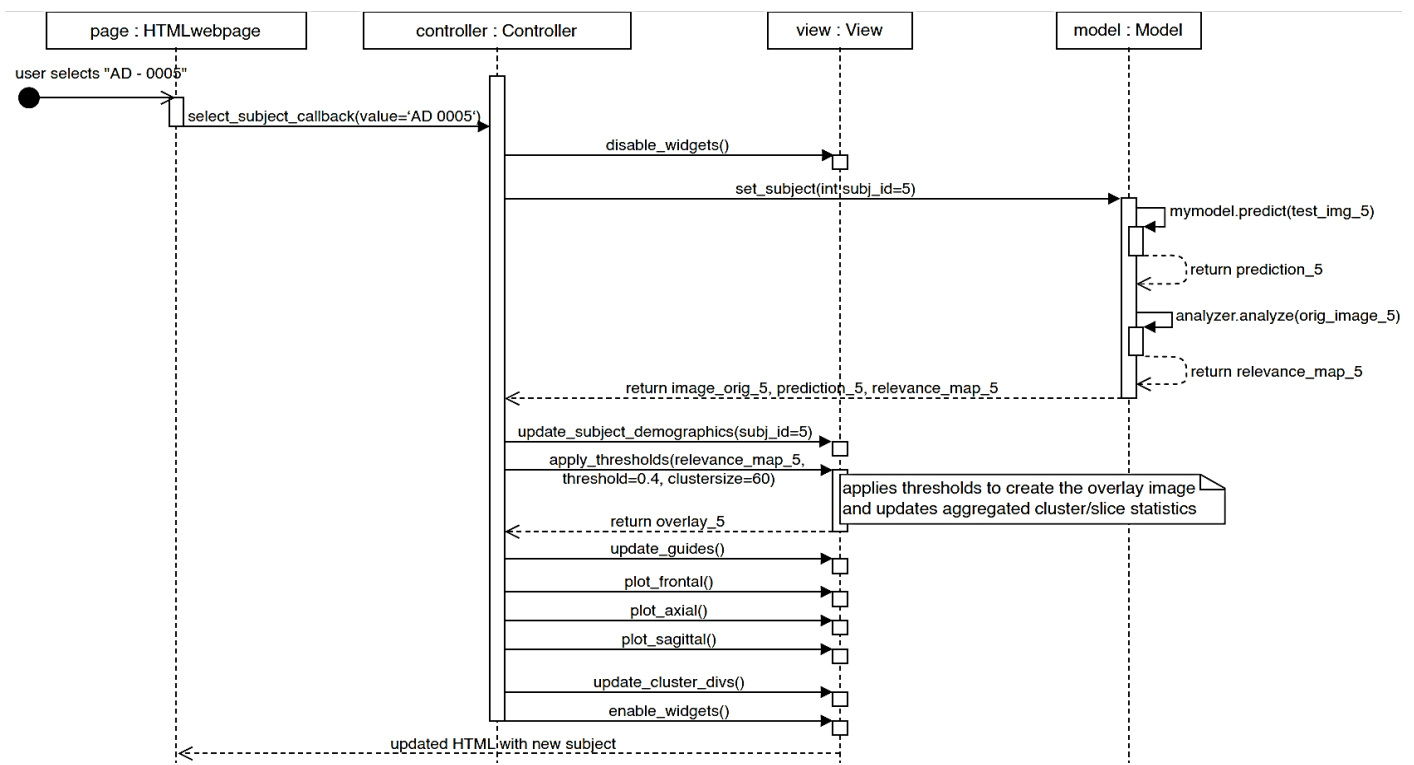

**Supplementary Figure 2**    Sequence diagram of function calls when selecting a new person.

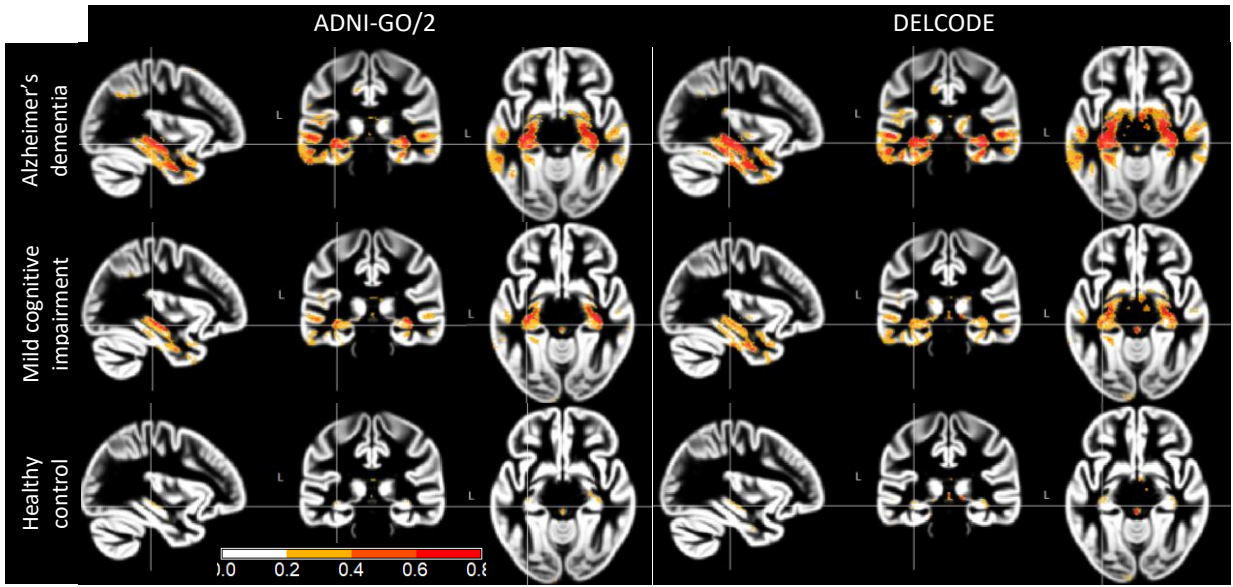

**Supplementary Figure 3** Comparison of mean relevance maps between samples. Left: ADNI-GO/2, Right: DELCODE.

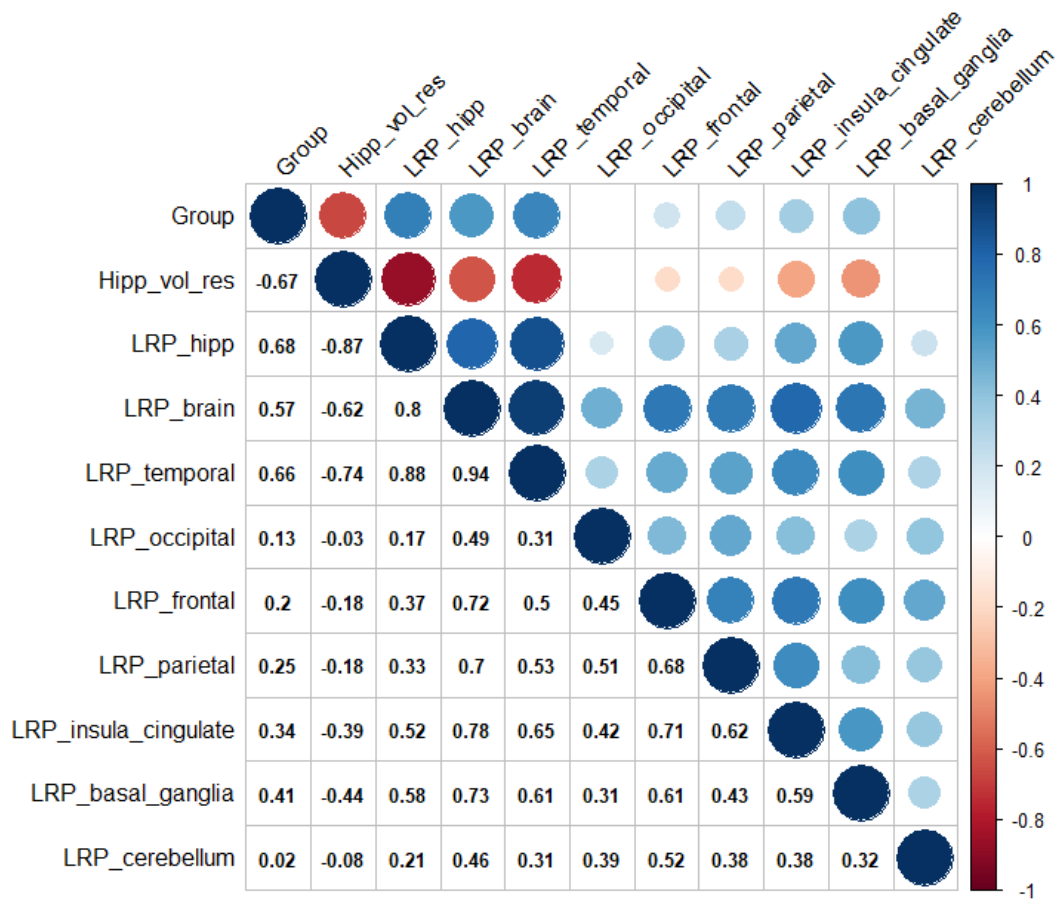

**Supplementary Figure 4** Correlation matrix of hippocampus volume (residualized) and several brain regions' relevance scores for DELCODE participants and the model trained on the whole ADNI-GO/2 dataset. The correlation between hippocampus volume and hippocampus relevance was highest (–0.87). Upper right triangle entries were thresholded a p<0.001. For simplicity, group was numerically encoded as CN=1, MCI=2, AD=3.